# A solvable quantum many-body problem

John Lekner

School of Chemical and Physical Sciences, Victoria University of Wellington, Wellington, New Zealand

john.lekner@vuw.ac.nz

**Abstract** The many-body wavefunction $e^{-KR}$, where $R^2$ is the sum of the squares of the interparticle distances $r_{ij}$, is shown to be the solution of Schrödinger's equation for a potential proportional to $R^{-1}$, for $N$ particles. A set of translationally invariant rotational eigenstates of angular momentum is found.

**Keywords**: many-body wavefunctions, exact solutions, rotational excited states

We consider a system of $N$ identical bosons, in its translationally invariant zero-momentum state. Let $R^2$ be the sum of the squares of the $\frac{1}{2}N(N-1)$ interparticle distances,

$$R^2 = \sum_{i=1}^N \sum_{j>i}^N \left(\boldsymbol{r}_i - \boldsymbol{r}_j\right)^2 = \sum_{i=1}^N \sum_{j>i}^N r_{ij}^2 = \frac{1}{2}\sum_{i=1}^N \sum_{j=1}^N r_{ij}^2 \qquad (1)$$

Connection with hyperspherical coordinates [1], in which the hyper-radius $r_h$ is defined by $r_h^2 = \sum_1^N r_i^2$, is via the identity $R^2 = Nr_h^2 - \{\sum_1^N \boldsymbol{r_i}\}^2$. $[R^2 = \frac{1}{2}\sum_{i=1}^N \sum_{j=1}^N \left(\boldsymbol{r}_i - \boldsymbol{r}_j\right)^2 = \frac{1}{2}\sum_{i=1}^N \sum_{j=1}^N \left\{r_i^2 + r_j^2 - 2\boldsymbol{r}_i \cdot \boldsymbol{r}_j\right\} = N\sum_1^N r_i^2 - \{\sum_1^N \boldsymbol{r_i}\}^2.]$

There is one known system for which Schrödinger's equation may be solved for any number of particles, namely that of pairwise harmonic interactions [2], with the potential function $V\left(\{r_{ij}\}\right) = \frac{1}{2}M\omega^2 R^2$. In three dimensions, harmonic interactions have the ground state energy and wavefunction

$$E = \frac{3}{2}(N-1)N^{1/2}\,\hbar\omega, \qquad \Psi = \exp\left\{-N^{-1/2}\frac{M\omega}{2\hbar}R^2\right\} \qquad (2)$$

Another solvable many-body problem is discussed here, with an interaction potential that is not pairwise except for $N = 2$, but is proportional to the reciprocal of the root-mean-square interparticle distance. It may be that the set of solutions give below can be useful in the modelling of helium micro- and nano-droplets [3-7].

We shall show that the many-body wavefunction $\Psi = e^{-KR}$ is an eigenstate of the Hamiltonian $H = -\frac{\hbar^2}{2M}\sum_1^N \nabla_i^2 + V(\{r_{ij}\})$, with the potential $V$ proportional to $R^{-1}$. In the two-body case this is of course just the familiar $r_{12}^{-1}$ Coulomb interaction. Allowing for the reduced mass $m = M/2$, the Schrödinger equation $H\Psi = E\Psi$ of the relative motion is solved by $\Psi_2 = e^{-Kr_{12}}, V_2 = -\frac{2\hbar^2 K}{M}\ r_{12}^{-1}$, with energy $E_2 = -\frac{\hbar^2 K^2}{M}$.

For three particles the wavefunction $\Psi = e^{-K(r_{12}+r_{13}+r_{23})}$ is only an approximate solution for a potential proportional to $r_{12}^{-1} + r_{13}^{-1} + r_{23}^{-1}$. However, $\Psi_3 = e^{-KR}$ with $R = \sqrt{r_{12}^2 + r_{13}^2 + r_{23}^2}$ is exactly an eigenstate of a potential proportional to $R^{-1}$. We can verify this in cartesian coordinates, with $H = -\frac{\hbar^2}{2M}\sum_1^N\left(\partial_{x_i}^2 + \partial_{y_i}^2 + \partial_{z_i}^2\right) + V(\{x_i, y_i, z_i\})$ in three dimensions. Alternatively, we can use relative coordinates $r_{ij}$ and operate on the proposed wave function with the three-body Laplacian for relative motion ($\partial_{12}$ stands for $\partial/\partial_{r_{12}}$, $c_{1,23} = \frac{r_{12}^2+r_{13}^2-r_{23}^2}{2r_{12}r_{13}}$, etc.)

$$\Lambda_3 = \partial_{12}^2 + \frac{2}{r_{12}}\partial_{12} + \partial_{13}^2 + \frac{2}{r_{13}}\partial_{13} + \partial_{23}^2 + \frac{2}{r_{23}}\partial_{23} + c_{1,23}\partial_{12}\partial_{13} + c_{2,13}\partial_{12}\partial_{23} + c_{3,12}\partial_{23}\partial_{13} \quad (3)$$

This kinetic energy operator is as in equation (10) of [8] or (31) of [9]. Again allowing for the pairwise reduced mass $m = M/2$, the Schrödinger equation becomes $E - V\left(\{r_{ij}\}\right) = -\frac{\hbar^2}{M}\Psi^{-1}\Lambda_3\Psi$, and is solved by $\Psi = e^{-K\sqrt{r_{12}^2+r_{13}^2+r_{23}^2}}$, $E_3 = -\frac{3\hbar^2 K^2}{2M}$, $V_3 = -\frac{15\hbar^2 K}{2M}\ R^{-1}$.

For $N = 4$ the relative motion Laplacian is (in agreement with equation (30) of [10])

$$\Lambda_4 = L_3 + \partial_{14}^2 + \frac{2}{r_{14}}\partial_{14} + \partial_{24}^2 + \frac{2}{r_{24}}\partial_{24} + \partial_{34}^2 + \frac{2}{r_{34}}\partial_{34} + c_{1,23}\partial_{12}\partial_{13} + c_{1,24}\partial_{12}\partial_{14} + c_{2,13}\partial_{12}\partial_{23} + c_{1,34}\partial_{13}\partial_{14} + c_{2,14}\partial_{12}\partial_{24} + c_{2,34}\partial_{23}\partial_{24} + c_{3,12}\partial_{13}\partial_{23} + c_{3,14}\partial_{13}\partial_{34} + c_{3,24}\partial_{23}\partial_{34} + c_{4,12}\partial_{14}\partial_{24} + c_{4,13}\partial_{14}\partial_{34} + c_{4,23}\partial_{24}\partial_{34} \quad (4)$$

With the Hamiltonian $H_4 = -\frac{\hbar^2}{M}\Lambda_4 + V_4$ and wavefunction $\Psi = e^{-KR}, R = \sqrt{r_{12}^2 + r_{13}^2 + r_{23}^2 + r_{14}^2 + r_{24}^2 + r_{34}^2}$ we find $V_4 = -\frac{16\hbar^2 K}{M} R^{-1}, E_4 = -\frac{2\hbar^2 K^2}{M}$.

The general result is the following: with $R^2 = \frac{1}{2}\sum_{i=1}^N \sum_{j=1}^N r_{ij}^2$, the wavefunction $\Psi = e^{-KR}$ is an eigenstate of the Hamiltonian in one dimensions for $N > 2$, and in two and three dimensions for $N \geq 2$, with energy equal to the number of particles times $\frac{\hbar^2 K^2}{2M}$, and potentials proportional to $R^{-1}$:

$$E_N = -N\frac{\hbar^2 K^2}{2M} \qquad \text{(one, two or three dimensions)}$$

$$V_N = -N(N-2)\frac{\hbar^2 K}{2MR} \qquad \text{(one dimension}, N > 2)$$

$$V_N = -N(2N-3)\frac{\hbar^2 K}{2MR} \qquad \text{(two dimensions)} \quad (5)$$

$$V_N = -N(3N-4)\frac{\hbar^2 K}{2MR} \qquad \text{(three dimensions)}$$

(The two-body case in one dimension needs special consideration: the wavefunction $e^{-K|x_1-x_2|}$ is the bound state of a delta function pair potential.) These results suggest that the energy associated with the wavefunction $e^{-KR}$ is always given by $E_N = -N\frac{\hbar^2 K^2}{2M}$, and the $D$-dimensional potential (with the exception for $D = 1, N = 2$ noted above) is

$$V_N = -N(DN - D - 1)\frac{\hbar^2 K}{2MR} \qquad (D \text{ dimensions}) \quad (6)$$

The binding energy for the $N$-body system is proportional to $NK^2$, while the potential energy strength is proportional to $N(3N-4)KR^{-1}$. To model a droplet of liquid helium 4 consisting of $N$ atoms, for example, we would keep $K$ constant. We are then replacing the physical short-range and approximately pairwise interactions with a single many-body potential proportional to $N^2KR^{-1}$ for large $N$. But note that $R$ is dominated by particle pairs with large separations, which is not physically reasonable for helium atoms.

*Rotational excited states.* (The following discussion is restricted to three dimensions.) The two-body states are just the hydrogenic wavefunctions, with reduced boson mass $M/2$ replacing the electron mass. The ground state is $e^{-KR}$ $(R = r_{12})$, the ground state energy $E_2^0 = -\frac{\hbar^2 K^2}{M}$ for potential $V_2 = -\frac{2\hbar^2 K}{M} R^{-1}$. An example of a translationally invariant excited state is $\Psi_2^{\mathrm{e}} = [x_1 - x_2 + i(y_1 - y_2)]e^{-K_e R}$. To keep the same interparticle potential we set $K_e = \frac{1}{2}K$. The energy is $-\frac{\hbar^2 K^2}{4M}$, one quarter of the ground state binding energy. The state $\Psi_2^{\mathrm{e}}$ is an eigenstate of the orbital angular momentum operator $L_z = L_{z_1} + L_{z_2} = i\hbar\{x_1\partial_{y_1} - y_1\partial_{x_1} + x_2\partial_{y_2} - y_2\partial_{x_2}\}$ with eigenvalue $\hbar$. Also $L_+ = L_{x_1} + iL_{y_1} + L_{x_2} + iL_{y_2}$ gives zero when operating on $\Psi_2^{\mathrm{e}}$, which has the uppermost $L_z$ eigenvalue of the angular momentum triplet.

For the three-body states with potential $V_3 = -\frac{15\hbar^2 K}{2M} R^{-1}$ $\left(R = \sqrt{r_{12}^2 + r_{13}^2 + r_{23}^2}\right)$ we have the ground state wavefunction $\Psi_3^0 = e^{-KR}$ and ground state energy $E_3^0 = -\frac{3\hbar^2 K^2}{2M}$. An excited state is, with $x_{12} = x_1 - x_2$ etc.,

$$\Psi_3^{\mathrm{e}} = [x_{12} + iy_{12}][x_{13} + iy_{13}][x_{23} + iy_{23}]e^{-K_e R} \tag{7}$$

This is a solution of Schrodinger's equation, with energy and potential

$$E_3^e = -\frac{3\hbar^2 K_e^2}{2M}, \qquad V_3^e = -\frac{33\hbar^2 K_e}{2MR} \tag{8}$$

To keep the potential equal to $V_3$ we set $K_e = \frac{5}{11}K$, which makes $E_3^e = \left(\frac{5}{11}\right)^2 E_3^0$. The uppermost state $\Psi_3^{\mathrm{e}}$ has eigenvalue $3\hbar$ of the angular momentum operator $L_z = \sum_{i=1}^3 L_{z_i}$.

A four-particle excited state of this form is $\Psi_4^{\mathrm{e}} = \prod_{i=1}^4 \prod_{j>i}^4 \left[x_{ij} + iy_{ij}\right] e^{-K_e R}$. Its energy and potential are $E_4^e = -\frac{2\hbar^2 K_e^2}{M}, V_4^e = -\frac{40\hbar^2 K_e}{MR}$. We equate $V_4^e$ to $V_4 = -\frac{16\hbar^2 K}{MR}$, which gives $K_e = \frac{2}{5}K$. The energy of the excited state is $E_4^e = \left(\frac{2}{5}\right)^2 E_4^0$. The angular momentum eigenvalue is 6ħ. For five particles we find $K_e = \frac{11}{31}K,\ E_5^e = \left(\frac{11}{31}\right)^2 E_5^0$. For this kind of excited state, for general $N$ in three dimensions,

$$\Psi_N^{\mathrm{e}} = \prod_{i=1}^{N}\prod_{j>i}^{N}\left[x_{ij} + iy_{ij}\right] e^{-K_e R}, \quad L_z\Psi_N^{\mathrm{e}} = \frac{1}{2}N(N-1)\Psi_N^{\mathrm{e}}, \quad L_+\Psi_N^{\mathrm{e}} = 0 \tag{9}$$

$$K_e = \frac{3N-4}{N^2+2N-4}K, \qquad E_N^e = \left(\frac{3N-4}{N^2+2N-4}\right)^2 E_N^0 = -\left(\frac{3N-4}{N^2+2N-4}\right)^2 N\frac{\hbar^2 K^2}{2M} \tag{10}$$

We note that, for fixed $K$ (and thus fixed binding energy per particle in the ground state) the excitation energy of these rotational eigenstates decreases with $N$, while the angular momentum increases quadratically with $N$.

**Acknowledgement** This paper has benefited from Reviewer comments and references.

**Competing interests** None to declare.

**Data availability** No data is associated with this paper.

**Funding** There was no funding.